\documentclass[ip,twocolumn]{jpsj3}

\usepackage{txfonts}
\usepackage{xcolor}
\usepackage{url}

\newcommand{\hto}{Ho$_{2}$Ti$_{2}$O$_{7}$}

\title{MIEZE Neutron Spin-Echo Spectroscopy of Strongly Correlated Electron Systems}

\author{Christian Franz$^1$\thanks{christian.franz@frm2.tum.de}, 
Steffen S\"aubert$^2$,
Andreas Wendl$^2$,
Franz X. Haslbeck$^{2,3}$,
Olaf Soltwedel$^{2,4}$,
Johanna K. Jochum$^1$,
Leonie Spitz$^2$,
Jonas Kindervater$^2$,
Andreas Bauer$^2$,
Peter B\"oni$^2$, and 
Christian Pfleiderer$^2$\thanks{christian.pfleiderer@tum.de}
}
\inst{$^1$Heinz Maier-Leibnitz Zentrum (MLZ), Technische Universit\"{a}t M\"{u}nchen, D-85748 Garching, Germany \\
$^2$Physik-Department, Technical University of Munich, D-85748 Garching, Germany \\
$^3$Institute for Advanced Study, Technical University of Munich, D-85748 Garching, Germany \\
$^4$Physik Department, Technical University of Darmstadt, G-64289 Darmstadt, Germany } %

\abst{Recent progress in neutron spin-echo spectroscopy by means of longitudinal Modulation of IntEnsity with Zero Effort (MIEZE) is reviewed. Key technical characteristics are summarized which highlight that the parameter range accessible in momentum and energy, as well as its limitations, are extremely well understood and controlled. Typical experimental data comprising quasi-elastic and inelastic scattering are presented, featuring magneto-elastic coupling and crystal field excitations in {\hto}, the skyrmion lattice to paramagnetic transition under applied magnetic field in MnSi, ferromagnetic criticality and spin waves in Fe. In addition bench marking studies of the molecular dynamics in H$_2$O are reported. Taken together, the advantages of MIEZE spectroscopy in studies at small and intermediate momentum transfers comprise an exceptionally wide dynamic range of over seven orders of magnitude, the capability to perform straight forward studies on depolarizing samples or under depolarizing sample environments, as well as on incoherently scattering materials. }

\begin{document}
\maketitle

\section{Motivation}

Neutrons may decipher vital microscopic and molecular mechanisms providing non-destructive and non-invasive structural and dynamic information. As the processes at mesoscopic and macroscopic scales are of particular interest, small angle neutron scattering has received great interest in a wide range of different topics. To unravel the dynamical properties in the regime of typical SANS experiments in many cases direct measurements of the intermediate scattering function are of central interest, as they permit direct comparison with theoretical simulations. Taken together with the need for a large dynamical range, this identifies neutron spin-echo spectroscopy that is optimized for small scattering angles as an exceptionally powerful method. 

Important examples of materials systems in which neutron spin-echo spectroscopy has proven to be extremely successful include polymers, reactive surfaces and interfaces, composite materials, macromolecular systems, biomolecules, catalytic interfaces, pharmaceuticals, fuel cells, gas storage materials, battery materials, thermoelectric materials, photovoltaics, and glass formation. In contrast, surprisingly few neutron spin-echo studies have been reported in archetypal hard condensed matter systems such as typical strongly correlated electron systems. Located at the cross-roads between pressing questions in the applied and fundamental sciences the same materials as tuned by temperature and/or non-thermal control parameters feature entirely different characteristics. 

Spectroscopic studies represent a primary tool for the exploration of the vast diversity of magnetic materials.  A prototypical example are second order phase transitions as a means to explore the validity of the present day understanding of critical phenomena on very large length and time scales\cite{1982Mezei,1986Farago,2017Kindervater}. A more recent example concerns the spin glass transition, where hidden processes in re-entrant spin glasses \cite{1981Gabay,2018Wagner,2018Saubert} and Berry phase contributions in chiral spin glasses have revived current interest \cite{2006Fabris,2010Campbell}. Last but not least, an important example concerns topological spin textures, such as skyrmion lattices in chiral magnets \cite{2009Muehlbauer,2013Janoschek}. All of these phenomena comprise microscopic dynamics touching on the limit of characteristic meso-scale textures, where very small energy scales dominate at large length-scales. 

On a related note, molecular magnets represent a new class of nano-scale magnetic materials based on multifunctional molecules aimed for spintronics as well as biological and medical applications. Namely, molecular magnets reveal highly unusual spin dynamics with distinct low-lying eigenstates that require careful experiments at low momentum transfers for identification. 

The origin of unconventional superconductivity in copper oxide-, rare earth- and iron-based materials and its relationship to magnetism continues to provide one of the great scientific enigmas of the 21st century. Many observations challenge the well-established BCS theory of phonon-driven superconductivity and help to benchmark theoretical models evoking magnetism. Great progress has been made in understanding the excitation spectra. Yet, studies at very high energy resolution will be required to distinguish between different scenarios. Moreover, the flux pinning and flux lattice melting are still fairly unexplored microscopically awaiting high resolution spectroscopic studies. 

Further, many batteries rely on ion exchange and/or the ionic conductivity of lithium and hydrogen. This makes them prime candidates for in-operando neutron scattering studies to visualize charge and discharge conditions. Quasi-elastic neutron scattering offers insights on diffusion processes on the weakly scattering lithium ions, probing dynamic properties, especially in lithium-ion battery materials. Also, gas storage materials such as hydrides, hydrates, clathrates, metal-organic frameworks and nanotubes are of great interest for reversible hydrogen storage, encapsulation of small molecules, such as alkanes, and trapping of harmful gases, such as carbon dioxide. Here measurements of the characteristic rotational spectrum of hydrogen models are uniquely able to identify the amount of un-bound hydrogen in a sample. Of similar interest is carbon dioxide sequestration and the uncontrolled release of methane and carbon dioxide from undersea clathrate beds. For improved studies, high-resolution neutron spectroscopy is required avoiding constraints imposed by the strong incoherent scattering of hydrogen. 

From a methodological point of view all of these topics are hard to address by conventional neutron spin-echo spectroscopy, as they either depolarize the neutron beam, exhibit their most interesting properties under depolarizing conditions, or generate strong incoherent scattering. Moreover, a multitude of coupled degrees of freedom play together, forming inherently complex interdependences covering a large parameter space that is exceedingly difficult to assess comprehensively. In particular the properties at small momentum transfers play an important role, which is basically not accessible with conventional neutron spectroscopy such as triple axis, time of flight, backscattering and even neutron spin echo.

The objective of this paper is it to illustrates the potential of so-called longitudinal neutron resonance spin-echo spectroscopy with specific emphasis on the MIEZE technique in studies of archetypal hard condensed matter problems. The paper begins with an introduction to the basic notions of neutron spin-echo spectroscopy Sec.\,\ref{notions}, revisiting historic prejudice against the MIEZE technique in Sec.\,\ref{revisited}. It then reviews major advances achieved in recent years with the MIEZE technique at the beam-line RESEDA \cite{Franz2015} at FRM II in Sec.\,\ref{RESEDA}. The progress made is, finally, illustrated in Sec.\,\ref{examples} in terms of selected examples that may also be viewed as bench-marker on a more general level.


\section{Basic notions of neutron spin-echo spectroscopy}
\label{notions}


Using conventional neutron scattering methods, such as triple-axis or time-of-flight spectroscopy, measurements at ultra-high energy resolution require a reduction of beam divergence and wavelength spread. In turn, this causes massive losses of beam intensity, which render real experimental studies impossible on a practical level. In comparison, neutron spin-echo spectroscopy achieves unprecedented energy resolution at high neutron beam intensities, as the energy transfer is encoded in the polarisation of the scattered neutron beam. This way, the energy resolution is effectively decoupled from the wave-length spread.

The key operational component in neutron spin-echo spectrometers are Larmor precession devices, where the effective field integral $J_0=\int B_0 \mathrm{d}L$ generated by these precession devices corresponds directly to the maximum resolution ($B_0$ and $L$ are the effective field strength and path length, respectively). Thus, the main factor limiting the resolution of neutron spin-echo spectrometers arises from differences of $J_0$ for individual neutron trajectories, causing a de-phasing of the neutron beam. 

\begin{figure*}[t]
\centering
\includegraphics[width=0.8\linewidth,clip=]{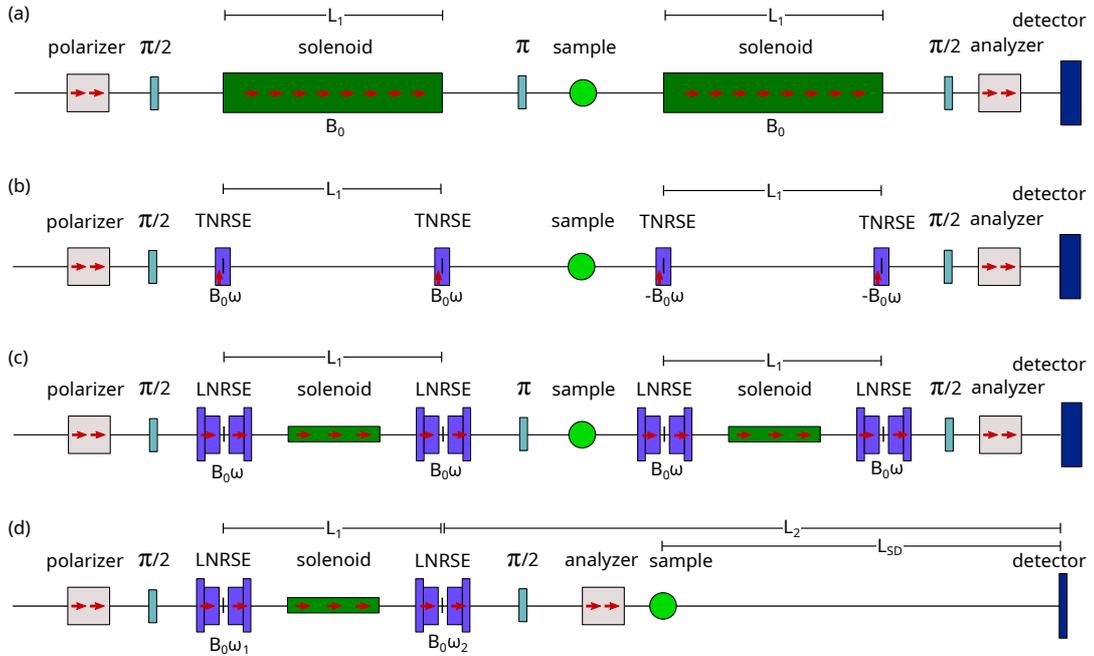}
\caption{(Color online) Different types of neutron spin-echo spectrometers. Red arrows indicate orientation of the field guiding the neutron polarization. (a) Conventional neutron spin-echo spectrometer (NSE). Long solenoids with DC fields $B_0$ act as precession devices (PDs). The precession is started and stopped by spin flippers. The flipper close to the sample inverts the effective polarity of $B_0$ of the first PD. (b) Resonance spin-echo spectrometer with $B_0$ transverse to the beam direction (TNRSE) and RF field $B_1$ transverse to $B_0$. A pair of RF spin flippers acts as PD. Guide fields parallel to $B_0$ can be applied, but are not necessary. The polarization is carefully coupled into the PD device. (c) Resonance spin-echo spectrometer with $B_0$ parallel (longitudinal) to the beam direction (LNRSE), $B_1$ is transverse to $B_0$. A small longitudinal guide field is applied along the beam path. (d) LNRSE in MIEZE configuration: The RF flippers are operated at different frequencies leading to a sinusoidal time modulation of the detector signal. The polarization analyzer is placed upstream the sample. Spin depolarizing samples or high magnetic fields in the sample region do not affect the modulated signal. Plot adapted from Ref.\,\citen{2016Krautloher}.}
\label{f1}
\end{figure*}

Three variants of neutron spin-echo spectrometers are established: (i) conventional Neutron Spin Echo (NSE) \cite{1980Mezei},  (ii) Neutron Resonance Spin Echo (NRSE) \cite{1987Golub,1996Koppe,2002Keller,2016Krautloher}, and (iii) Modulation of IntEnsity with Zero Effort (MIEZE) \cite{1992Gahler,2006Bleuel,2016Krautloher}.  Depicted in Fig.\ref{f1}\,(a) are the basic principles of a conventional NSE spectrometer as originally proposed by Mezei \cite{1972Mezei}. Here the precession devices represent large solenoids that generate a static magnetic field $B_0$. The NSE set-up is characterized by differences of field integral for both parallel and divergent neutron trajectories.  As the magnetic field strength for trajectories parallel to the central axis of a solenoid varies with the distance to the central axis squared, typical variations of field integrals are of the order $\delta J/J\approx 10^{-3}$ for standard neutron beam cross-sections of a few $\mathrm{cm}^2$. Compensation of these differences of neutron paths require carefully designed correction coils, reducing the variations of the field integral to values as low as $\delta J/J\approx 10^{-6}$ at state of the art beam-lines such as IN11 \cite{IN11_web} or IN15 \cite{2015Farago}. 

In comparison to NSE the Larmor labelling in NRSE is implemented by pairs of compact radio frequency spin flippers (RF flippers) marking the boundaries of the precession regime as shown in Fig.\,\ref{f1}\,(b) and \ref{f1}\,(c). Depicted in Fig.\ref{f1}\,(b), is the implementation of NRSE historically realized first. Here, the static field, $B_0$, and the radio-frequency field , $B_{\rm rf}$, used in the resonant spin-flippers are transverse, i.e., perpendicular, to the neutron beam. This configuration is also referred to as transverse NRSE (TNRSE). As a main disadvantage of TNRSE, transmission losses limit the cross section of the coils in the spin flippers used for the static fields, and thus the energy resolution. Due to these difficulties and further challenges when correcting path length differences, TNRSE spectrometers never reached the same high energy resolution achieved by conventional NSE spectrometers, especially for small momentum transfers. Nonetheless, TNRSE features several important advantages as compared to NSE, notably it permits to use Larmor labelling for high-resolution triple-axis spectroscopy \cite{2015Keller} and high-resolution diffraction \cite{1999Rekveldt, 2002Keller_ApplPhysA}.

An alternative NRSE design that allows to overcome the limitations of transverse NRSE in quasi-elastic scattering is the so-called longitudinal NRSE (LNRSE) shown Fig.\ref{f1}\,(c) \cite{2005Haussler}. For full details we refer to Ref.\,\citen{2016Krautloher} and references therein. In comparison to TNRSE, the static field of the RF spin flippers is here  longitudinal (parallel) to the neutron beam (the RF field is unchanged perpendicular to the neutron beam). While the longitudinal orientation of the $B_0$ field implies differences of field integral for individual trajectories akin those encountered in conventional NSE, these deviations cancel for inhomogeneities that are symmetric with respect to the central axis, because the RF spin flip in the center of the DC coil entails a sign change of the Larmor phase. In comparison to NSE, in LNRSE typical inhomogeneities of the field integral are as low as $\delta J/J_0 \sim 3 \times 10^{-4}$, i.e., they are at least a factor of three smaller. Moreover, the same correction coils used in NSE (Fresnel or Pythagoras) allow to reduce $\delta J/J_0$ further. Thus existing technology allows, in principle, to achieve an energy resolution a factor of three larger than for existing NSE beam-lines.

Moreover, as the physical dimensions of the coils used for Larmor labelling in LNRSE are rather compact, a performance equivalent to NSE may be expected at much reduced electric power densities as well as much reduced costs for manufacturing. Indeed, for the longitudinal field geometry the polarization can be guided adiabatically through the instrument such that magnetic shielding becomes dispensable in comparison to TNRSE. Further, the Larmor labelling as implemented by RF spin flippers allows for an elegant field integral subtraction \cite{2005Haussler,2016Krautloher}, which avoids changes of the set-up as encountered with so-called shorties in NSE. In fact, the field integral subtraction used in LNRSE permits to increase the dynamic range of the spectrometer substantially towards very small spin-echo times ($\ll$ps) as compared with NSE without need for any changes of the instrument. 

The NRSE concept, finally, allows to implement the MIEZE method, as schematically depicted in Fig.\ref{f1}\,(d). Operating two RF spin flippers in the primary spectrometer at different frequencies, a sinusoidal time-dependence of the beam intensity is generated at the detector when the MIEZE condition 
\begin{equation}
\frac{L_1}{L_2}=\frac{\omega_2-\omega_1}{\omega_1}
\end{equation}
is satisfied. The energy transfers is thereby encoded in the contrast of the intensity modulation, where the spin-echo (or MIEZE) time, $\tau_{\mathrm{MIEZE}}$, is given by
\begin{equation}
\tau_{\mathrm{MIEZE}}=\frac{m^2}{h^2}\lambda^3 \Delta \omega L_{\mathrm{SD}}
\end{equation}
$m$ is the mass of the neutron, $h$ is Planck's constant, $\lambda$ is the neutron wave-length and $ L_{\mathrm{SD}}$ is the distance between sample and detector. It is helpful to note that $ L_{\mathrm{SD}}$ quantitatively compares with the length of typical NSE solenoids (at RESEDA L$_{\mathrm{SD}}$\,=\,1.87\,m). As the Larmor labelling is entirely realized in front of the sample the polarization analyzer may also be placed in front of the sample, resulting in two major advantages. First, the signal contrast is insensitive to depolarizing samples and sample environment, e.g., ferromagnetic and superconducting materials may be investigated and studies under strong magnetic fields may be carried out. Second, polarization analysis can be readily implemented. Third, the signal contrast becomes insensitive to incoherent scattering, facilitating studies of hydrogen-containing materials.


\section{Assessment of MIEZE spectroscopy revisited}
\label{revisited}

It has long been appreciated that the MIEZE method may be particularly helpful in studies of dynamic processes in the parameter range of typical SANS experiments \cite{2005Bleuel,2005Bleuel2}. Accordingly, construction of dedicated MIEZE beam-lines had been proposed before, e.g., at the ORNL-SNS. The technical concerns, expressed as a result of the assessment of these proposals, were based on the experience with transverse RF spin flippers available at the time. However, in recent years great progress has been achieved with LNRSE and especially longitudinal MIEZE as reviewed below. In the light of this progress, it is instructive to revisit the concerns raised historically against dedicated MIEZE instruments as they appear to reflect presently accepted views in the scientific community. 

First, it has been argued that spin-echo spectroscopy at small momentum transfers may be carried out at conventional NSE beam-lines. However, conventional NSE spectrometers cannot easily be adapted to the needs of small angle scattering. Namely, NSE spectrometers typically have no capabilities for beam collimation in the primary spectrometer arm. They are optimized for large samples covering the entire beam cross-section ($\sim40 \times 40\,\mathrm{mm}^2$). The situation may be improved, in principle, as illustrated by recent efforts to perform NSE-reflectometry studies at J-NSE \cite{2014Holderer}. Yet, the spatial resolution of the detectors at conventional NSE spectrometers are limited and essentially insufficient for the small momentum transfers of real SANS geometries. Moreover, the beam-paths include sections in air, and the correction elements (Fresnel/Pythagoras coils) are comparatively thick both generating considerable small-angle scattering background.

Second, it has been shown that NSE studies on depolarizing samples are possible by virtue of the so-called ferromagnetic spin-echo technique\cite{1985Boucher,1986Farago}. This appears to suggest that the capabilities of the MIEZE technique to perform studies on depolarizing samples and sample environments are not unique. However, the need for additional polarizers before and after the sample in ferromagnetic NSE arrives at the expense of losses of intensity by at least a factor of four, where the mutual influence between the various magnetic fields may causes problems. Considerable background is also generated in a small-angle geometry due to the additional components in the beam. The intensity penalty for this somewhat acrobatic information recovery is therefore rather high, as emphasized in Ref.\citen{1986Farago}.

Third, to reach an energy resolution equivalent to conventional NSE spectrometers RF spin flips have to be performed at frequencies as high as several MHz. In comparison, for transverse RF spin flippers the concomitant difficulties to correct variations of the field integral, $\delta J/J_0$, limit the maximum driving frequencies for spin flips to a few hundred kHz. It is therefore believed that the necessary high flipping frequencies cannot be reached. Using a bootstrap configuration (in TNRSE) increases the effective Lamor frequency (f$_{\textrm{eff}}$) of the neutron by a total factor of four (as compared to classical NSE), placing the maximum Lamor frequency reached for NRSE-TAS at 3.2\,MHz \cite{2002Mezei} and for TMIEZE at 1\,MHz \cite{2007Hayashida}. Employing longitudinal RF spin flipper resonant spin flips up to 1\,MHz (f$_{\textrm{eff}}$ = 2\,MHz) are by now in regular operation. Moreover, proof-of-concept studies have demonstrated resonant spin flips up to 3.6\,MHz (f$_{\textrm{eff}}$ = 7.2\,MHz), clearly establishing the feasibility to reach 8\,MHz (f$_{\textrm{eff}}$ = 16\,MHz).

Fourth, it is frequently argued that Fourier times as high as 100\,ns may only be reached for tiny scattering angles, as the detector would need a spatial depth resolution of less than 0.05\,mm. Therefore, the maximum timescales accessible in MIEZE are believed to be generically limited to below 1\,ns. However, the recent measurements of the neutron phase front across the entire detector area and thus detector waviness demonstrates unambiguously that spin-echo signals exceeding 100\,ns may be recorded under perfectly controlled conditions. It is even technically feasible to correct the waviness of individual detector foils. Moreover, specification of current GEM technology promises further improvements of the temporal and spatial resolution needed for studies approaching 1000\,ns\,\cite{nGem}. 

Fifth, it is argued that for inelastic scattering processes NSE offers substantial advantages as compared to MIEZE. Notably, the availability of polarization analysis at NSE enables routinely studies of inelastic spin incoherent scattering, e.g., to single out incoherent scattering background on hydrogen atoms. Yet, the standard setup MIEZE is sensitive to all fluctuations in a sample. If polarisation analysis is needed, MIEZE provides a polarized beam at the sample position by design. With an additional analyser after the sample not only a spin flip/ non-spin flip analysis is straight-forward. In fact, MIEZE may also be combined with a full 3D polarisation analysis using a miniaturized device for spherical polarization analysis\,\cite{2014Kindervater}. 

Sixth, it is believed that placing the analyzer in front of the sample introduces a sensitivity to differences of flight path that limits the scattering angles geometrically such that even for small sample cross-sections the detector depth becomes a critical issue. However, the geometry effects have been simulated by means of various methods\,\cite{2013Weber,2011Brandl, 2018Martin}. Experimental studies are in excellent agreement with these calculations as summarized below. All studies clearly show that scattering angles as encountered in typical SANS geometries are highly favorable for the MIEZE technique. Indeed, flight path differences of neutron trajectories from the sample to the detector play only a minor role and the full resolution may be exploited without restrictions on the sample geometry\,\cite{2018Martin}. 

In the light of these considerations, which suggest that MIEZE is particularly amenable for small angle scattering geometries, it is also helpful to comment on possible routes towards dedicated MIEZE-SANS beam-lines.  Notably, to the best of our knowledge all existing SANS beam-lines offering polarized neutrons use transverse guide fields. This is inherently incompatible with the requirements of a longitudinal MIEZE. Moreover, a modular upgrade of existing SANS beam-lines is not favorable, as the studies performed so far demonstrate the need for very accurate alignment implemented best by means of permanent installations. In addition, existing SANS beam-lines typically face considerable overload factors, suggesting that additional MIEZE options will aggravate the demand for regular SANS beam time.

A promising route towards high resolutions provides also the combination of TOF and MIEZE at a spallation source. In contrast to conventional MIEZE, small deviations from the MIEZE condition do not result in a loss of contrast, but in small frequency deviations that may be used to tune the instrument fast and efficiently. The development of TOF-MIEZE is pursued at the LARMOR spectrometer at ISIS and within the VIN ROSE spin echo instrument suite at J-PARC MLF, where the MIEZE option is available for users \cite{2016Oda, 2017Hino, 2017Nakajima}.

Last but not least, the development of compact MIEZE modules\,\cite{2011Georgii} offer an interesting add-on option for many existing beam-lines. However, the dynamical range accessible with existing modular MIEZE devices vis a vis the MIEZE condition mentioned above are massively constrained by the reduced dimension of these modules. In other words, use of a MIEZE module would be on the expense of a highly unfavorable MIEZE ratio as compared with the potential parameter range accessible at a dedicated MIEZE beam-line.


\begin{figure*}[t]
\centering
\includegraphics[width=0.95\linewidth,clip=]{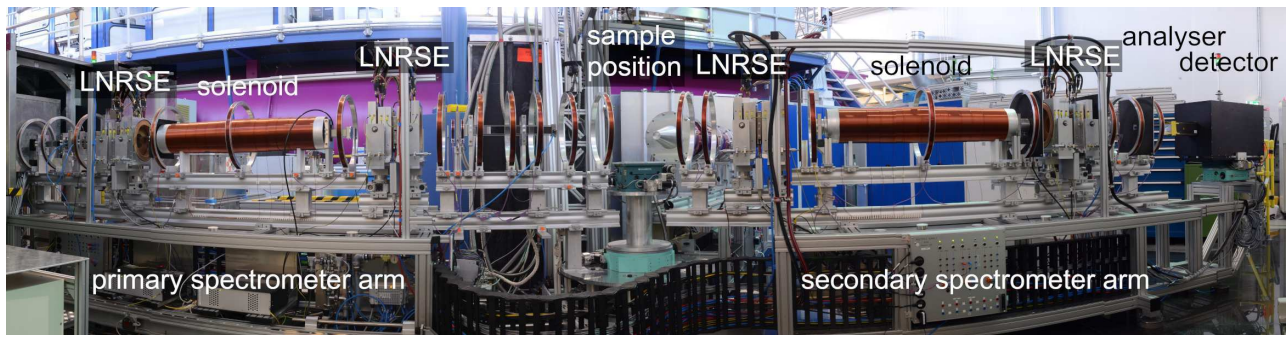}
\caption{(Color online) Neutron spin-echo spectrometer RESEDA at FRM II. The complete reconstruction of the beam-line since 2015 permitted to demonstrate the full potential of LNRSE as summarized below. The configuration shown here corresponds to the set-up depicted in Fig.\,\ref{f1}\,(c). When operated as a MIEZE spectrometer the secondary spectrometer arm supports a flight tube and the analyser is placed in front of the sample position.}
\label{f2}
\end{figure*}

\section{Resonance spin-echo spectrometer RESEDA}
\label{RESEDA}

Recent advances with longitudinal NRSE have been achieved at the beam-line RESEDA at FRM II. In this section the configurations of RESEDA following the reconstruction of the instrument since 2015 are summarized. In addition, the main tests establishing the accessible parameter range vis a vis the considerations mentioned in Sec.\,\ref{revisited} in longitudinal MIEZE are reviewed.

\subsection{Current instrument configuration}

RESEDA (REsonance Spin Echo for Diverse Applications) is a high-resolution resonance spin-echo spectrometer located at the cold neutron guide NL5-S in the Neutron Guide Hall West at FRM II. In its present configuration the instrument provides access to a large  range of spin-echo times and scattering vectors for quasi-elastic measurements. A recent photograph of RESEDA is shown in Fig.\,\ref{f2}. It supports in particular longitudinal neutron resonance spin-echo (LNRSE) in the range from below 0.001\,ps to above 20\,ns for $\lambda = 10\,{\rm \AA}$ \cite{2003Haussler,2016Krautloher} and modulation of intensity with zero-effort (MIEZE) in the range from below 0.001ps to above 10 ns for $\lambda = 10\,{\rm \AA}$) \cite{2018Franz,2018Wendl,Wendl-preprint}. A standard $^3$He detector and a 2D CASCADE detector with an active area of $20 \times 20\,\mathrm{cm}^2$ are available. The CASCADE detector is characterized by a spatial resolution of $2.4\times2.4\,\mathrm{mm}^2$ and a temporal resolution up to 100\,ns\,\cite{nGem,2014Kindervater}.

In its present status RESEDA makes use of a clean and well-defined polarized neutron beam at wave-lengths between 3.5\,{\AA} and 15\,{\AA} (at larger wavelengths the intensity is massively reduced and suitable for proof-of-concept studies only). The wavelength spread may be varied between 7.7\,\% and 17.2\,\%. Resonant spin flips for the longitudinal configuration may be routinely performed at frequencies up to 1.2\,MHz. An optimized solenoid permits to reduce the effective field integral, $J_0$, such that a dynamic range of over seven orders of magnitude between a few Hz up to above 1\,MHz is available. Indeed, in the MIEZE configuration an amplitude modulation as low as 0.2\,Hz could be demonstrated. The corresponding effective spin-echo time extends from below a few 100\,fs to above 100\,ns, covering over seven orders of magnitude in dynamic range. 

Originally conceived and constructed as a transverse NRSE instrument, RESEDA was converted into a longitudinal NRSE starting in 2015. The initial instrument design of RESEDA as a transverse NRSE dates back to the late 1990s, when it was proposed by Bleuel as part of his PhD thesis\cite{2003Bleuel}. Bleuel also set up the first version of the instrument, many components of which could only be replaced in recent years for lack of resources. At the time RESEDA was proposed, the NRSE technique represented a new idea with little or no practical experience. In turn, starting from the RF spin flippers a large number of the components had to be developed alongside the basic instrument concept.

The commissioning of RESEDA as NRSE instrument began with the operation of FRM II in 2004. 
In 2006 it was realized that due to activation of Co, approximately 40\,m of the polarizing neutron guide had to be replaced by a non-polarizing guide combined with a polarizing cavity resulting in a long down-time and a new realignment of the beamline.  Following this, corrosion problems in the RF coils and a loss of beam polarisation due to slightly magnetized mu-metal components of the RF coils caused further delays. Due to a large number of technical problems and severe understaffing, it was not before 2011 when first scientific studies could be attempted. Until then some of the most important challenges included the design of the RF spin flippers, the design and installation of a first polarizer, the development of first detector electronics, design and installation of a velocity selector and the implementation of a first version of the instrument software.  

Starting around 2011, the transverse MIEZE option could be commissioned, making straight-forward measurements under applied magnetic fields possible for the first time. Major constraints during this time remained the unreliable mechanical set up and the lack of instrument status polling, resulting in typical tuning times of the beam-line in excess of several weeks, as well as major problems with the CASCADE detector (electronic failure and a slow drift), which required substantial efforts to be resolved. Moreover, during this period two implosions of the neutron guide NL5 caused longer down-times. By the end of 2013 a first test of the LNRSE technique proved to be extremely successful, clearly demonstrating that essentially all of the technical limitations encountered with TNRSE in quasi-elastic scattering experiments could be resolved. 

\begin{figure*}[t]
\centering
\includegraphics[width=0.95\linewidth]{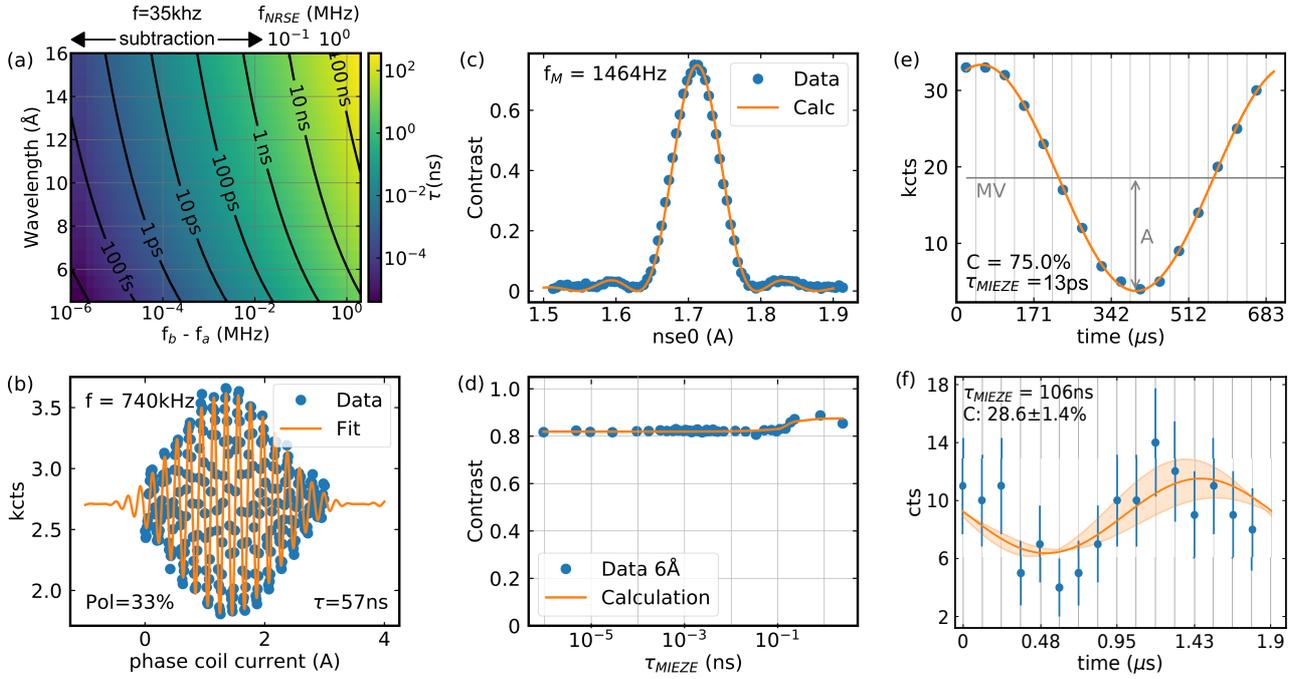}
\caption{(Color online) Parameter range accessible with the MIEZE set-up and typical experimental data. (a) Currently accessible spin-echo times as a function of modulation frequency, covering over seven orders of magnitude in dynamic range. (b) Characteristic LNRSE spin-echo signal at $\tau=57\,\mathrm{ns}$ as recorded at RESEDA. (c) Calibration scan across the the MIEZE condition; this scan is equivalent to the observation of the spin-echo in LNRSE. (d) Variation of the MIEZE contrast as a function of the dynamically accessible range. Data are in excellent agreement with theoretical expectation (line). The reduced value at small spin-echo times and the maximum at 1 ns are due to the Bloch-Siegert shift. (e) MIEZE contrast at a low modulation frequency of 1.464\,kHz.  (f) MIEZE signal at a spin-echo time of 106\,ns as recorded at a wavelength of 20\,{\AA} with a large polarization of nearly 30\,\%. }
\label{f3}
\end{figure*}

The reconstruction of RESEDA as a dedicated LNRSE beam-line started in 2015. It commenced by rebuilding the spectrometer from scratch, beginning with a rigid, modular platform of the primary and secondary spectrometer arms, as well as a complete reconstruction of the electrical wiring and the instrument control system. Following this a large number of major changes were implemented comprising:
\begin{itemize}
\item Replacement of the polarizing cavity in the incident beam by a double V-cavity polarizer to remove parasitic signal contamination and to increase the accessible maximum wavelengths from 10 to 15\,{\AA}.
\item Replacement of the polarizing cavity before the sample by a transmission bender analyser for the MIEZE set-up to remove parasitic signal contamination.
\item Development and installation of new coils for optimal control of the guide fields.
\item Development and installation of a new field integral subtraction coil to reduce beam inhomogeneities, permitting field integral subtraction down to spin-echo times $\tau\approx0$.
\item Development and implementation of new RF-circuits, including new amplifiers, impedance matching devices, a bespoke automated capacitance box, RF-coils and readout via oszilloscope, for resonant spin flips up to 3.6\,MHz.
\item Construction and installation of neutron guides on the primary spectrometer arm, resulting in a flux increase exceeding a factor of five.
\item Construction and implementation of Mezei-type $\pi/2$ flippers replacing inherently unstable spin flippers causing substantial operational instabilities.
\item Design and implementation of a MIEZE flight tube for background reduction.
\item Design and implementation of an automated crossed-slit system for controlled beam collimation.
\item Development and implementation of spectrometer software based on the NICOS standard at FRM II. 
\item Implementation of full status polling of all instrument components as part of the new spectrometer software.
\end{itemize}

The changes implemented at RESEDA as part of its reconstruction prove to be essential for the parameter range described above. Perhaps most impressive, however, is the reproducibility and speed at which different configurations may be reached. This is best illustrated by the tuning time for the beam-line, which could be reduced from two weeks to less than one hour. This underscores, that LNRSE and LMIEZE despite its inherent complexity may be operated and performed routinely.

\begin{figure*}[t]
\centering
\includegraphics[width=0.8\linewidth]{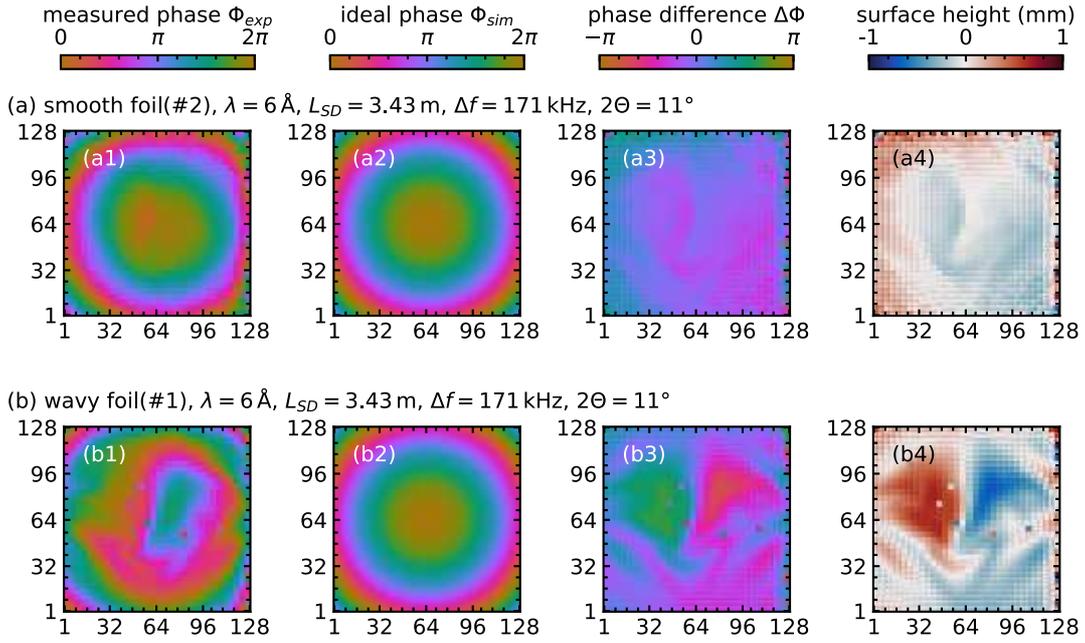}
\caption{(Color online) Variation of the neutron phase as recorded for a single detector foil of the CASCADE detector at RESEDA and comparison with the expected ideal behavior to infer the foil waviness. Panels (a1) through (a4) display the properties of a smooth foil (No.\,2); panels (b1) through (b4) display the properties of a wavy foil (No.\,1). (a1) and (b1) Measured phase on the detector, $\Phi_{\mathrm{exp}}$. (a2) and (b2) Expected ideal phase without waviness, $\Phi_{\mathrm{sim}}$. (a3) and (b3) Phase difference between measured and expected behaviour, $\Delta\Phi$. (a4) and (b4) Variation of surface height inferred from $\Delta\Phi$. Note the accuracy of determination of the foil warping of better 0.01\,mm.}
\label{f5}
\end{figure*} 

\subsection{Present limitations}

As the NRSE technique is technically very demanding, involving the manipulation of polarized neutrons and the concomitant interference of different spin states, it is instructive to address in further detail the limitations of the parameter regime currently accessible. In fact, the level of control and understanding of these limitations underscores the excellent control that has been achieved in recent years.

Shown in Fig.\,\ref{f3} is a compilation of key information on the accessible range of spin-echo times at RESEDA. Fig.\,\ref{f3}\,(a) displays the spin-echo times as a function of effective modulation frequency for LNRSE or MIEZE and neutron wavelength (left hand axis). The lower $x$-axis represents the frequency of intensity modulation in MIEZE spectroscopy. The upper $x$-axis represents the LNRSE frequency. For the right hand side of the upper $x$-axis the field integral is adjusted starting from modulation frequency of 35\,kHz. The plot covers the regime of wavelengths between 3.5\,{\AA} and 15\,{\AA} presently accessible at RESEDA. It is important to note the exceptional dynamic range from below 100\,fs up to above 100\,ns. 

A typical LNRSE spin-echo signal recorded at RESEDA for a large spin-echo time of $\tau=57\,\mathrm{ns}$ is shown in Fig.\,\ref{f3}\,(b). Tuning the difference of the Larmor phase of the spin-up and the spin-down states by means of the current of the phase coil, the characteristic interference between the two wave packages is observed. The large polarization of 33\,\% underscores the high quality of the signal. An increase of the polarization is anticipated upon better alignment of the RF-spin flippers, technically not possible at the time of this test. The same tests and theoretical simulations revealed in addition that the polarization was also limited by the inhomogeneities of the $B_0$ field, which is currently generated by two bespoke Helmholtz pairs. These inhomogeneities may be much reduced, when generating the $B_0$-field with optimized small superconducting solenoids, thereby further improving the overall performance.

A MIEZE tuning scan of the phase difference between the spin-up and spin-down wave-packages as tracked by the signal contrast of the intensity modulation is shown in Fig.\,\ref{f3}\,(c). This scan corresponds to the spin-echo signal observed in LNRSE as shown in Fig.\,\ref{f3}\,(b). At the maximum of the signal contrast the MIEZE condition is satisfied. It is important to emphasize the excellent agreement with theoretical expectation notably the peak value, peak width and presence of faint side maxima \cite{2019Jochum}.  

\begin{figure*}[t]
\centering
\includegraphics[width=0.95\linewidth]{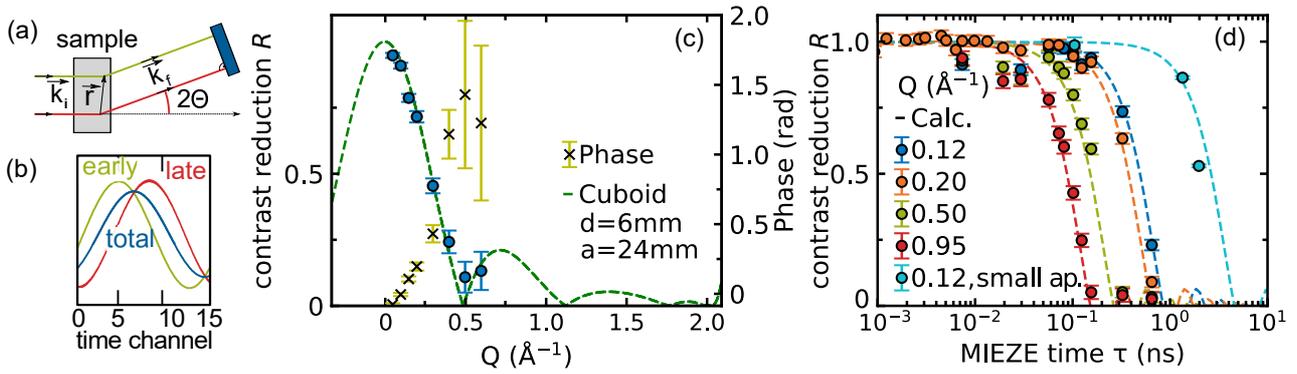}
\caption{(Color online) Characterization and bench marking of signal reduction due to sample geometry. (a) Schematic depiction of the path length differences. (b) Differences of the intensity modulation as recorded at the time-resolved detector for different trajectories. (c) Calibration of the geometrical contrast reduction by means of graphite with a diameter of 24\,mm. Excellent agreement with theoretical calculation is observed (green line). The phase of the signal at the detector is shifted reflecting which trajectories dominate. (d) Reduction of the MIEZE contrast for different momentum transfers in a disc-shaped powder of {\hto} with a diameter of 13\,mm (thickness 3\,mm), in excellent agreement with theoretical prediction. For larger wavelengths the reduction is shifted to larger spin-echo times (not shown), underscoring the feasibility of MIEZE at very large spin-echo times.}
\label{f4}
\end{figure*}

In general, the maximum energy loss in neutron scattering is given by the energy of the incident neutrons, which for cold neutrons is typically in the range 1\,meV to 5\,meV. In contrast, the energy transfer on the energy gain side may be much larger as, for example, in time-of-flight spectroscopy. As emphasized above longitudinal NRSE and longitudinal MIEZE are able to overcome the limits of classical NSE for high energy transfers by means of an effective field integral subtraction method\,\cite{2005Haussler}. Namely, a conventional NSE solenoid between the RF spin flipper allows to reduce the field integral without risk of depolarizing the beam. This way, in principle, infinitesimally small Fourier times may be reached, while the beam polarisation is retained. In turn, LNRSE and LMIEZE are able to cover an exceptionally large dynamic range from meV to neV energy transfers, elegantly encompassing several different spectroscopic techniques, notably time-of-flight, backscattering, and classical spin echo. We return a more detailed discussion of the data treatment and interpretation at these very short spin-echo times below.

Shown in Fig.\,\ref{f3}\,(d) is the variation of the maximum value of the contrast as a function of MIEZE times $\tau_{\rm MIEZE}$ for the entire dynamic range of over seven orders of magnitude. The contrast is in excellent agreement with calculations taking into account the correct treatment of the Bloch-Siegert shift. In particular the change of value around 0.1\,ns may be accurately reproduced \cite{2018Franz}.

With the current combination of coils used in the RF spin flippers the maximum frequency of 1.2\,MHz is due to the capacitive stray coupling between the $B_0$- and RF-coils. Use of a small superconducting solenoid in combination with the RF coil permitted to reduce the capacitive stray coupling, such that RF spin flips could be demonstrated up to 3.6 MHz, the maximum frequency accessible with the existing electronic circuits and RF generators. These tests establish, that the use of superconducting solenoids will allow to reach spin flip frequencies up to 8\,MHz, representing the anticipated frequency limit of the CASCADE detector currently available \cite{2011Haeussler,2016Koehli}.

A typical time dependence of the MIEZE signal is shown in Fig.\,\ref{f3}\,(e) for the low modulation frequency of 1.464\,kHz of the tuning scan shown in Fig.\,\ref{f3}\,(c). Using the amplitude $A$ and the mean value $MV$ the contrast of the signal $C$ may be determined. Convoluting the MIEZE contrast $C$ as determined for different MIEZE times with the resolution function shown in Fig.\,\ref{f3}\,(d), the intermediate scattering function $S(q,\tau)$ is obtained (several examples are presented in the next section).

Similarly, for the MIEZE set-up a signal contrast of nearly 30\,\% of the direct beam could be recorded at a record high spin-echo time of $\tau = 106\,{\rm ns}$ as shown in Fig.\,\ref{f3}(f). The latter was recorded for a neutron wave-length of $\lambda = 20.7\,{\rm \AA}$, at which the intensity is no longer sufficient for real experimental studies. Increasing the highest frequency of the RF spin flipper to 8\,MHz will allow to reach the same MIEZE spin-echo time of $\tau = 106\,{\rm  ns}$ at a wavelength of $\lambda = 10\,{\rm \AA}$, i.e., for a beam intensity that is larger by well over three orders of magnitude. Moreover, for the present configuration of RESEDA the record high MIEZE spin-echo time of $\tau = 106\,{\rm ns}$  could only be measured for the direct beam due to the background at small momentum transfers. The latter strongly motivates dedicated efforts to improve the SANS background at RESEDA in the future.

The calibration data shown in Fig.\,\ref{f3}\,(d) is shown for MIEZE times up to 3\,ns. For large MIEZE times the contrast observed experimentally decreases, but is still as high as $\sim 30\%$ at $\sim 100\mathrm{ns}$. Careful measurements of the neutron phase front, presented in Fig.\,\ref{f5}, allow to trace this reduction to the waviness of the detector foils. Namely, as part of the reconstruction of RESEDA accurate determination of the neutron phase was installed. The capability of phase-locked signal detection permits to spatially resolve the precise phase of the neutrons on the detector. As illustrated in Figs.\,\ref{f5}(a) and \ref{f5}(b) the six detection foils in the CASCADE detector presently available at RESEDA exhibit differences of waviness. In turn, the waviness may be corrected numerically. This demonstrates, that meaningful data may be collected at spin-echo times beyond $100\,\mathrm{ns}$. Improvements of the foil waviness in the next generation CASCASDE detector promise to improve the overall performance significantly.

The detection of the phase front on the detector (concentric rings) illustrates also the importance of the angular dependence of the path length differences. This raises the question for signal degradation due to path length difference between different locations at the sample as illustrated in Fig.\,\ref{f4}(a) and \ref{f4}(b). A calibration of the geometrical contrast reduction of the MIEZE signal by means of a graphite sample with a diameter of 24\,mm as a function of momentum transfer $q$ is shown in Fig.\,\ref{f4}(c). For the typical angular range of SANS studies the reduction at momentum transfers up to $\sim0.3\,\mathrm{\AA}^{-1}$ is moderate and accompanied by a weak increase of the averaged signal phase. Excellent agreement of the reduction factor with theoretical calculation is observed (green line). 

The understanding of the effects of geometrical signal degradation has also been illustrated in a study of {\hto} powder in a disc-shaped sample holder with a diameter of 13\,mm (thickness 3\,mm) as shown in Fig.\,\ref{f4}(d). For different momentum transfers nearly perfect agreement with theoretical prediction is observed. For larger wavelengths and tighter collimation the reduction is shifted to larger spin-echo times, underscoring the feasibility of MIEZE at very large spin-echo times.

\begin{figure*}[t]
\centering
\includegraphics[width=0.8\linewidth]{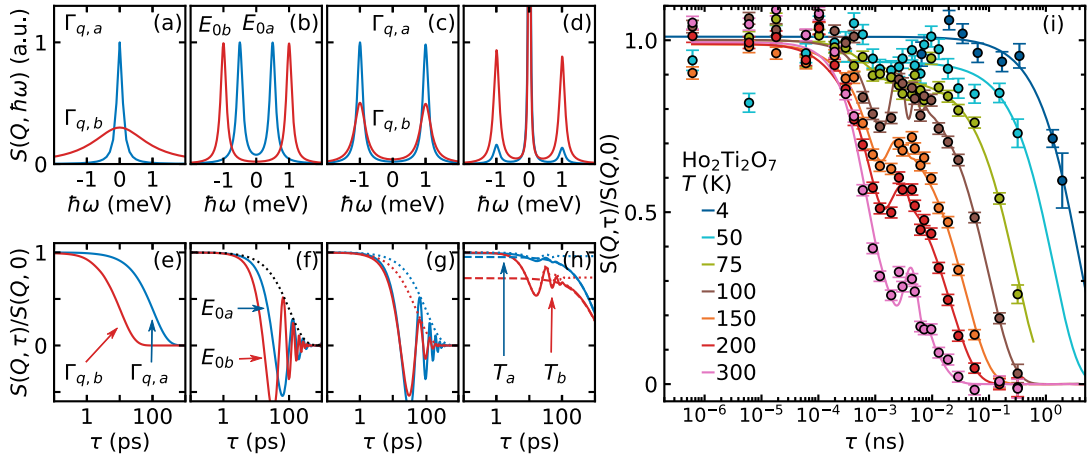}
\caption{(Color online) Typical MIEZE data expected for quasi-elastic and inelastic excitations. (a) through (d): Typical scattering functions $S(q,\tau)$ expected of quasi-elastic and inelastic excitations (red and blue denote different temperatures). (e) through (h) Typical intermediate scattering functions $S(q,\tau)$ of the excitations shown in panels (a) through (d).(i) Experimental data recorded in high quality powder of {\hto} at different temperatures. Quasi-elastic scattering due to paramagnetic fluctuations and crystal field excitations are clearly resolved. Lines represent fits corresponding to the processes illustrated in panels (a) through (h), providing accurate values of the excitation energy and damping.}
\label{f6}
\end{figure*}

\section{Bench marking experimental studies}
\label{examples}

In this section the potential of longitudinal MIEZE is illustrated by means of selected examples featuring both strong fluctuations and low-lying inelastic excitations. First, data recorded in the geometrically frustrated rare earth pyrochlore system {\hto} are reviewed, representing a widely established bench marker for quasi-elastic scattering. This is followed by data on the emergence of skyrmion lattice order in MnSi as an example for measurements under magnetic field. Revisiting seminal work performed by means of so-called ferromagnetic NSE, the third example concerns the observation of spin waves in the vicinity of the Curie temperature of Fe. Last but not least, first results on the dynamical properties of pure water are reported, illustrating the advantages of the MIEZE technique in materials generating strong incoherent scattering.

\subsection{Paramagnetic fluctuations and crystal field excitations}

The capability to detect accurate phase-resolved information on the CASCADE image detector  as implemented at RESEDA enables measurements of quasi-elastic and inelastic processes. An illustration of such a combination of quasi-elastic processes and an inelastic (damped) excitation is presented in Fig.\,\ref{f6}. The well-known typical energy and momentum dependence as observed in the scattering function $S(q,\omega)$ is shown in Figs.\,\ref{f6}(a) through \ref{f6}(d). Blue and red lines refer to differences of characteristic damping and excitation energy as expected at different temperatures. The corresponding intermediate scattering functions $S(q,\tau)$ are shown in Figs.\,\ref{f6}(e) through \ref{f6}(h), where the panels on equivalent excitation energies and damping parameters are placed underneath each other. It is helpful to note the quantitative energy and time scales. In particular, whereas the quasi-elastic response, shown in Fig.\,\ref{f6}(e), may be well known, the combined signal of the quasi-elastic and inelastic signal, shown in Fig.\,\ref{f6}(h), may be less familiar.

The capability to detect combined quasi-elastic and inelastic scattering over an exceptionally large dynamic range has been studied at RESEDA in selected pyrochlore oxides, A$_2$B$_2$O$_7$. Characterized by rare earth magnetic moments located on the A site of a three- dimensional network of corner sharing tetrahedra these materials are model systems of geometric frustration, where {\hto} is well known for the emergence of spin ice properties at low temperatures\cite{2001Bramwell,2009Fennell,2012Krey}. Instrumental for the spin ice state are the crystal field levels, which impose a strong local Ising anisotropy of the rare earth spins along the $\langle111\rangle$ direction in the cubic unit cell. In turn, the rare earth moments at low temperatures are either pointing into or out-of the tetrahedra. An unresolved scientific question concerns the putative collective character of the spin excitations as well as the presence of magneto-elastic coupling with the crystal lattice. The latter may provide a constraint of the spin ice state that lifts the degeneracy in the zero temperature limit.

At high temperatures {\hto} displays strong thermal fluctuations of the rare earth ions. As the moments are large and the effects of magnetic anisotropies weak, these fluctuations have become an important bench marker for quasi-elastic neutron scattering instrumentation. Seminal studies using conventional NSE have observed an intermediate scattering function characteristic of momentum independent quasielastic scattering at spin-echo times in the range $\tau\approx 4\,\mathrm{ps}$ up to 2\,ns at temperatures between 0.2\,K and 200\,K \cite{2004Ehlers}. While it was speculated that a slight reduction of $S(q,\tau)$ below $4\,\mathrm{ps}$ may provide putative hints of an inelastic process, the precise nature of this process could not be determined. Further insights on this inelastic process were observed in recent time of flight (ToF) and backscattering measurements, suggesting the presence of an Orbach process at intermediate temperatures in which a phonon drives a crystal field excitation\cite{2009Clancy,2017Ruminy}. In addition, the observation of additional crystal electric field transitions was reported. In contrast, below 50\,K these studies lacked the necessary resolution to resolve the slow spin dynamics. 

Taken together, the information observed in NSE, ToF and backscattering highlights the need for data recorded across the entire parameter range as recorded in a single measurement. On the one hand, this allows to track the evolution of different features, frequently located at the boundaries of the individual parameter regimes of the different types of spectrometers. On the other hand, this provides complete information on $S(q,\tau)$ without need to Fourier transform part of the data for a comprehensive account. This promises to avoid ambiguities and greatly increases the speed at which a full data set may be obtained.

Presented in Fig.\,\ref{f6}(i) are the intermediate scattering functions of a high-quality powdered sample of {\hto} as measured at RESEDA at intermediate and small momentum transfers. It is helpful to note that the dynamic range covers over seven orders of magnitude. For the parameter range in which data had been reported in the literature excellent quantitative agreement is observed. Further, measurements down to very short spin-echo times extend deep into the regime, where the conventional spin-echo approximation assuming a linear relationship between Larmor phase and energy transfer is no longer valid. To shed further insights on the data recorded at these very short spin-echo times, the full response of the spectrometer was simulated by different methods and found to be tractable and physically meaningful \cite{2018Franz, 2018Wendl}. A full account of the study on {\hto} including information on the interpretation at very short spin-echo times is beyond the scope of this review and will be presented elsewhere \cite{Wendl-preprint}. 

\begin{figure}
\centering
\includegraphics[width=0.95\linewidth]{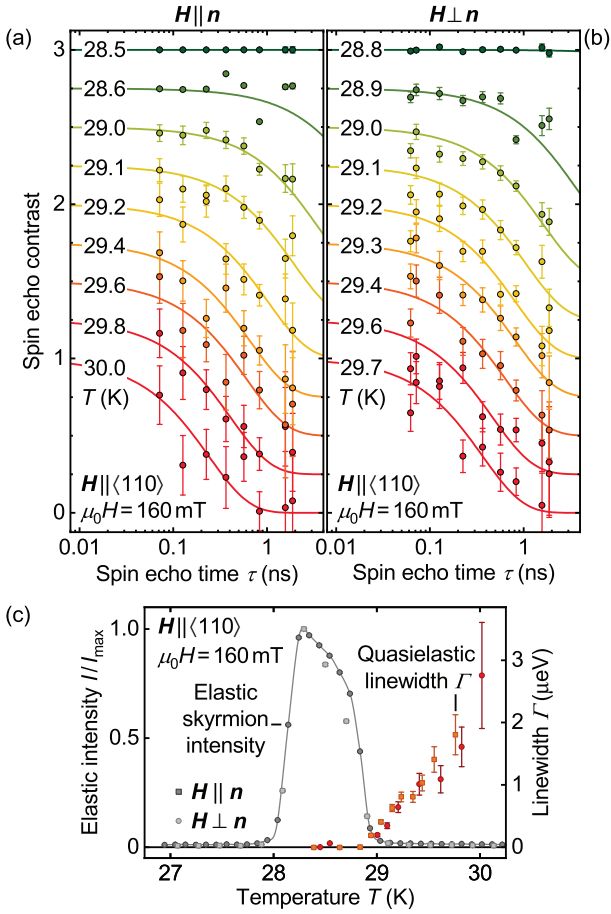}
\caption{(Color online) Evolution of spin fluctuations in MnSi across the phase transition between the skyrmion lattice and paramagnetism. (a) Typical intermediate scattering functions at various temperatures for field parallel to the neutron beam. (b) Typical intermediate scattering functions at various temperatures for field perpendicular to the neutron beam. (c) Comparison of the temperature dependence of the elastic scattering intensity associated with the skymrion lattice and quasielastic linewidth $\Gamma$.}
\label{f7}
\end{figure}

\subsection{Depolarizing sample environments: magnetic fields}

Conventional neutron spin-echo spectroscopy requires excellent control of the polarization of the neutron beam along the entire beam trajectory up to the analyzer in front of the detector. Use of depolarizing sample environments, such as large magnetic fields or depolarizing sample containments, are therefore prohibitively difficult. Even ferromagnetic NSE does not permit straight-forward measurements under large applied magnetic fields, as the stray field of the sample magnet will interfere with the additional polarizers and analyzers required in the conventional NSE set up.

In contrast, already early proof of concept measurements demonstrated that the MIEZE contrast is unaffected by magnetic fields up to 17\,T, the maximum available at the time. In fact, four magnet systems have been tested at RESEDA so far: (i) a bespoke set of uncompensated water-cooled Helmholtz pairs for magnetic fields up to several hundred mT, (ii) an actively shielded 5T split-pair superconducting magnet system \cite{muhlbauer}, that is optimized for small angle neutron scattering, (iii) an uncompensated 17\,T solenoid in which the beam is oriented along the cylinder axis \cite{2012Holmes}, and, (iv) an uncompensated 2\,T split pair magnet system using high-$T_c$ superconductors. 

A large number of scientific questions may be clarified by neutron spectroscopy with ultra-high energy resolution under large magnetic fields. These include not only magnetic phase transitions and superconducting textures, but also soft matter and biological systems in which large magnetic fields may be instrumental for the creation of contrast and textures\cite{2013Liebi}.

In the following the importance of neutron spin-echo spectroscopy under applied magnetic fields will be illustrated for the case of the skyrmion lattice phase of the B20 compound MnSi. Forming a kind of vortex lines parallel to an applied magnetic field, the skyrmion lattice in MnSi is inherently two dimensional in symmetry. Since their initial identification in MnSi, an abundance of materials systems have been discovered featuring skyrmions and skyrmion lattice order. Despite this huge range of activities an unresolved pressing question concerns the precise kinetics of formation at the paramagnetic to skyrmion lattice transition at high temperatures. In view of the strongly two dimensional character of the skyrmion lattice the transition kinetics parallel and perpendicular to the applied magnetic field are of interest.

Shown in Figs.\,\ref{f7}\,(a) and \ref{f7}\,(b) are typical intermediate scattering functions $S(q,\tau)$ of MnSi recorded by means of the longitudinal MIEZE set-up at RESEDA. Data were recorded under a comparatively small applied magnetic field of 160\,mT for various selected temperatures across the paramagnetic to skyrmion lattice transition. Both the behaviour for field parallel and perpendicular to the direction of the incident neutron beam were investigated, where essentially identical behaviour was observed.

A comparison of the linewidth inferred from the MIEZE data as compared with the integrated elastic intensity observed in small angle neutron scattering is shown in Fig.\,\ref{f7}\,(c). With increasing temperature the skymrion lattice phase stabilizes in a well-defined temperature range. As the elastic scattering intensity vanishes with increasing temperature, the linewidth increases monotonically. Perhaps most remarkable, even though the differences of topological character of the skyrmion lattice phase and the paramagnetic suggest a first order transition consistent with the abrupt drop of the elastic intensity, there are no hints for a discontinuous increase of $\Gamma$ as might be expected. Assessment of the nature of the melting process at the skyrmion lattice to paramagnetic transition and the underlying spin fluctuations requires further experimental probes, notably the magnetization, ac susceptibility, specific heat as well as high resolution small angle neutron scattering and further spectroscopic probes. 

\begin{figure}[t]
\centering
\includegraphics[width=0.95\linewidth]{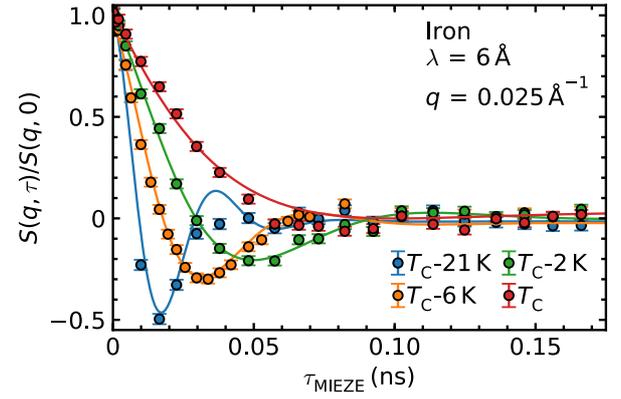}
\caption{(Color online) MIEZE signal in the ferromagnetic state of Fe. Ferromagnetic spin waves may be resolved on the background of strong paramagnetic fluctuations (note the linear time scale)\cite{2018Saubert,2019_Saeubert_PRB}.}
\label{f9}
\end{figure}

\subsection{Depolarizing materials}

Conventional NSE spectroscopy of depolarizing materials such as ferromagnets or superconductors is just as prohibitively difficult as measurements for depolarizing sample environments addressed in the previous section. One of the first proof of principle studies demonstrating the possibility to adapt conventional NSE studies such that depolarizing materials could be investigated concerned the spin waves just below the Curie temperature in Fe \cite{1982Mezei,1986Farago}. The specific adaptation of conventional NSE to depolarizing materials is also known as ferromagnetic spin-echo \cite{1982Mezei,1986Farago}.

Representing one of the archetypical ferromagnets, a recent MIEZE study at RESEDA pursuit the critical dynamics at the transition from the para- to the ferromagnetic phase in iron \cite{Kindervater-PhD,2017Kindervater,2018Saubert}. As a follow-up on this investigation recent measurements of the spectrum of spin waves in the ferromagnetic phase in iron are reviewed. Representing a strong scatterer, MIEZE measurements in Fe under typical SANS geometries proved to be straight forward, directly revealing the presence of spin waves with very high resolution as shown in Fig.\,\ref{f9} \cite{2018Saubert,2019_Saeubert_PRB}.

With increasing temperature and decreasing momentum the spin wave energy decreases. Further, with increasing momentum the energy of the spin waves increases when going from the zone centre to the boundary of the Brillouin zone. Closer inspection reveals clear deviations from the behaviour expected of a classical Heisenberg ferromagnet, where the spin wave energy varies quadratically with momentum. Due to the large magnetic moment of iron, it is necessary to take into account dipolar interactions in order to describe the observed spin wave energy \cite{1940Holstein,1966Keffer}.

The temperature dependence of the spin wave stiffness fits a power law dependence with a critical exponent $\mu\,=\,0.35\pm0.01$ consistent with theory. The critical exponent fits the data within a standard deviation of $\pm\sigma$, including all measured temperatures between $T_{\mathrm{C}}-1\,$K and $T_{\mathrm{C}}-21\,$K in the fit. The temperature dependence of the spin wave stiffnesses $D$ is in excellent agreement with the literature \cite{1969Collins,2018Saubert,2019_Saeubert_PRB}. All of these studies underscore the potential of MIEZE as a simple technique for very high-resolution spectroscopy of depolarizing samples and depolarizing sample environments, without the need for the instrumental complexities faced in ferromagnetic spin echo.

\subsection{Incoherently scattering materials}

A severe limitation of neutron scattering studies of hydrogen-based systems concerns the presence of incoherent scattering. Comprising two thirds spin-flip and one third non-spin-flip processes, incoherent scattering forcibly generates a background in all neutron scattering methods based on scattering of polarized neutrons, notably conventional NSE. At the same time soft matter systems are amongst the most intensely studied class of materials in which quasi-elastic neutron spectroscopy plays a vital role as the physical processes typically cover a very wide dynamic range down to very slow time-scales. Moreover, in recent years an increasing number of hydrogen-containing hard condensed matter systems has generated great interest. Important examples include energy-relevant materials as well as novel magnetic and superconducting compounds. In turn, neutron spectroscopic methods that can handle incoherent scattering are of basic interest to a large community.

The perhaps scientifically most important hydrogen-based system, which may serve as an important bench marker from a methodological point of view, represents pure water at ambient conditions. A large body of experimental and theoretical studies address the static and dynamical properties of water.  At the heart of the amazing complexities of water is the hydrogen bond (HB), which adds a supramolecular length scale that affects the structural relaxation of the liquid at the molecular level, i.e., the molecule diffusivity, and the time scale of reorientational dynamics in a hydrogen bond network. At the same time, scattering of hydrogen in water directly reflects the extreme sensitivity of neutrons to hydrogen due to the exceptionally large incoherent scattering cross section. In particular, cold neutrons with wavelengths of a few {\AA} and energies of several meV allow investigations of molecular motions on a nano- to picosecond time scale. 

\begin{figure}[t]
\centering
\includegraphics[width=0.95\linewidth]{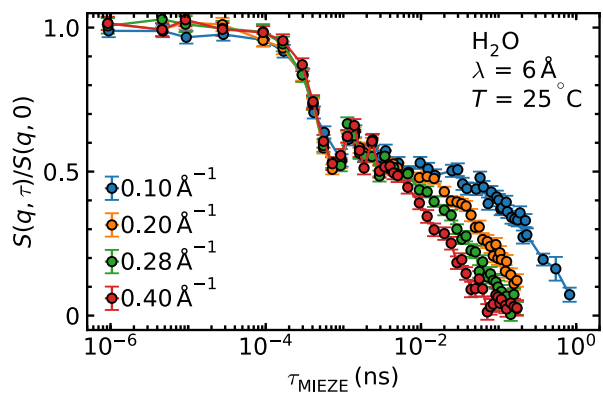}
\caption{(Color online) Typical MIEZE data recorded in H$_2$O. Data provide evidence of at least three dynamical processes, comprising transverse diffusion, transverse diffusion and liberation. The analysis of the spectra require full account of resolution effects beyond the review reported here and will be presented elsewhere.\cite{Soltwedel-preprint}}
\label{f8}
\end{figure}

Shown in Fig.\,\ref{f8} is the intermediate scattering function of pure water at ambient conditions as recorded with the MIEZE technique at RESEDA. In comparison to conventional NSE, which is limited to time scales above 4\,ps, seven orders of magnitude of dynamic range range are covered here. Data recorded under equivalent conditions at JNSE for $\tau>4\,{\mathrm{ps}}$ agree very well with the MIEZE data, when taking into account the ambiguities of the calibration procedure of the NSE signal (data and details of the comparison will be presented elsewhere\cite{Soltwedel-preprint}). 

For spin-echo times below the limit of conventional NSE the MIEZE data provides unambiguous evidence of a dynamical process with a characteristic time-scale around 0.3\,ps. This observation is consistent with a proposal based on neutron backscattering spectroscopy and dielectric spectroscopy \cite{2016:Arbe:PRL}. Moreover, revisiting preliminary neutron ToF data provide also putative evidence of such a process (cf. enhanced scattering intensity at 7.2 meV in Fig.16, p C3.22 of Ref.\,\cite{2012:Monkenbusch:IFF}). 

Taken together, the MIEZE data shown in Fig.\,\ref{f8} provide for the first time direct evidence that the molecular dynamics, at the heart of the exceptional properties of water, comprises at least three mechanisms in only one spectra. First, consistent with the literature, we find evidence for bulk transverse diffusion on length scales between 15 and 60\,\AA. As the main result, the exceptionally large dynamic range of the MIEZE technique from 0.01 up to 1000\,ps, reveals at least two more processes on sub-picosecond time-scales. Most likely, this represents direct evidence of vibration and liberation of water molecules. Here the incoherent one phonon like inelastic scattering is expected to resemble integration over all $q$ as a function of energy, which is proportional to the vibration density of states. The third process appears to be $q$-independent and occurs on time scale close to the expected life time of hydrogen bonds proposed by Arbe {et al.} \cite{2016:Arbe:PRL}. This highlights the potential of MIEZE spectroscopy as a unique method to access the long-sought information of the origin of the anomalous properties of water and related systems.

\section{Conclusions}

In conclusion, we have reviewed recent advances in resonant neutron spin-echo spectroscopy as a novel method for the investigation of low-lying spin fluctuations and inelastic excitations in strongly correlated electron systems. Focussing on the MIEZE technique based on longitudinal resonant spin flips, a well-understood and controlled exceptionally large dynamic range of over seven orders of magnitude are available for studies at low-momentum transfers typically covered in elastic small angle neutron scattering measurements. 

The opportunities associated with the advances in instrumentation described in this review are advertised in terms of selected bench-marking studies of quasi-elastic and inelastic processes, featuring magneto-elastic coupling and crystal field excitations in {\hto}, the skyrmion lattice to paramagnetic transition under applied magnetic field in MnSi, ferromagnetic criticality and spin waves in Fe, and molecular dynamics in H$_2$O. Further scientific highlights not covered in this paper for lack of space include the helimagnetic to paramagnetic quantum phase transition in cubic chiral magnets\,\cite{2014Kindervater,Kindervater-PhD}, frozen spin dynamics in the re-entrant spin glass system Fe$_{1-x}$Cr$_x$ \cite{2018Saubert}, and the cross-over between dynamical properties characteristic of local-moment and itinerant-electron magnetism in the superconducting ferromagnet UGe$_2$ \cite{2019_Haslbeck_PRB}. These and further studies currently under way illustrate a wide range of scientific progress by means of resonant neutron spin-echo spectroscopy in strongly correlated electron systems.

We wish to thank Andrew Boothroyd, Claudio Castelnovo, Prabakh Dharmalingam, Georg Ehlers, Marc Janoschek, Thomas Keller, and J\"urgen Neuhaus for discussions and support as well as generously sharing their insights. Financial support through DFG TRR80, SPP 2137 (Skyrmionics), DFG-GACR WI3320, ERC advanced grant TOPFIT, and BMBF (RESEDA-Plus: 05K16WO6) and the Munich Centre for Quantum Science and Technology (MCQST) are gratefully acknowledged. This project has received funding from the European Research Council (ERC) under the European Union´s Horizon 2020 research and innovation programme (grant agreement No 788031). The {\hto} samples were kindly supplied by Prabakh Dharmalingam and Andrew Boothroyd as funded by EPSRC grant EP/N034872/1. 

\bibliographystyle{jpsj}

\begin{thebibliography}{}

\end{thebibliography}


\begin{thebibliography}{10}

\bibitem{1982Mezei}
F.~Mezei: Phys. Rev. Lett. {\bfseries 49} (1982) 1096.

\bibitem{1986Farago}
B.~{Farago} and F.~{Mezei}: Physica B+C {\bfseries 136} (1986) 100.

\bibitem{2017Kindervater}
J.~Kindervater, S.~S\"aubert, and P.~B\"oni: Physical Review B {\bfseries 95}
  (2017) 014429.

\bibitem{1981Gabay}
M.~Gabay and G.~Toulouse: Phys. Rev. Lett. {\bfseries 47} (1981) 201.

\bibitem{2018Wagner}
J.~N. Wagner, W.~H\"au\ss{}ler, O.~Holderer, A.~Bauer, S.~M. Shapiro, and
  P.~B\"oni: Quantum Beam Science {\bfseries 2} (2018) 26.

\bibitem{2018Saubert}
S.~S\"aubert: {{PhD Thesis}}, Technische Universit\"at M\"unchen (TUM) (2018).

\bibitem{2006Fabris}
F.~W. Fabris, P.~Pureur, J.~Schaf, V.~N. Vieira, and I.~A. Campbell: Phys. Rev.
  B {\bfseries 74} (2006) 214201.

\bibitem{2010Campbell}
I.~A.~Campbell and D.~C.~M. C.~Petit: Journal of the Physical Society of Japan
  {\bfseries 79} (2010) 011006.

\bibitem{2009Muehlbauer}
S.~M{\"u}hlbauer, B.~Binz, F.~Jonietz, C.~Pfleiderer, A.~Rosch, A.~Neubauer,
  R.~Georgii, and P.~B{\"o}ni: Science {\bfseries 323} (2009) 915.

\bibitem{2013Janoschek}
M.~Janoschek, M.~Garst, A.~Bauer, P.~Krautscheid, R.~Georgii, P.~B\"oni, and
  C.~Pfleiderer: Phys. Rev. B {\bfseries 87} (2013) 134407.

\bibitem{Franz2015}
C.~Franz and T.~Schr{\"{o}}der: Journal of large-scale research facilities
  JLSRF {\bfseries 1} (2015) A14.

\bibitem{1980Mezei}
F.~Mezei, The principles of neutron spin echo, Neutron Spin Echo. Lecture Notes
  in Physics, Vol. 128. Springer, Berlin, Heidelberg, 1980.

\bibitem{1987Golub}
R.~Golub and R.~G{\"a}hler: physics letters A {\bfseries 123} (1987) 43.

\bibitem{1996Koppe}
M.~K\"oppe, P.~Hank, J.~Wuttke, W.~Petry, R.~G\"ahler, and R.~Kahn: Journal of
  Neutron Research {\bfseries 4} (1996) 261.

\bibitem{2002Keller}
T.~Keller, R.~Golub, and R.~G{\"ah}ler, Neutron Spin Echo -- A Technique for
  High-Resolution Neutron Scattering, In R.~Pike and P.~Sabatier (eds), {\em
  Scattering}. Academic Press, San Diego, 2002.

\bibitem{2016Krautloher}
M.~Krautloher, J.~Kindervater, T.~Keller, and W.~H{\"a}ussler: Review of
  Scientific Instruments {\bfseries 87} (2016) 125110.

\bibitem{1992Gahler}
R.~G\"ahler, R.~Golub, and T.~Keller: Physica B: Condensed Matter {\bfseries
  180-181} (1992) 899.

\bibitem{2006Bleuel}
M.~Bleuel, M.~Br\"oll, E.~Lang, K.~Littrell, R.~G\"ahler, and J.~Lal: Physica
  B: Condensed Matter {\bfseries 371} (2006) 297.

\bibitem{1972Mezei}
F.~Mezei: Zeitschrift f{\"u}r Physik A Hadrons and nuclei {\bfseries 255}
  (1972) 146.

\bibitem{IN11_web}
Information at ILL website:
  https://www.ill.eu/users/instruments/instruments-list/in11/characteristics/.

\bibitem{2015Farago}
B.~Farago, P.~Falus, I.~Hoffmann, M.~Gradzielski, F.~Thomas, and C.~Gomez:
  Neutron News {\bfseries 26} (2015) 15.

\bibitem{2015Keller}
T.~Keller and B.~Keimer: Journal of large-scale research facilities JLSRF
  {\bfseries 1} (2015) 37.

\bibitem{1999Rekveldt}
M.~T.~H. Rekveldt and W.~H. Kraan: Journal of Neutron Research {\bfseries 8}
  (1999) 53.

\bibitem{2002Keller_ApplPhysA}
T.~Keller, M.~Rekveldt, and K.~Habicht: Applied Physics A {\bfseries 74} (2002)
  s127.

\bibitem{2005Haussler}
W.~H{\"a}ussler and U.~Schmidt: Phys. Chem. Chem. Phys. {\bfseries 7} (2005)
  1245.

\bibitem{2005Bleuel}
M.~{Bleuel}, L.~B. {Lurio}, K.~{Littrell}, R.~{Gaehler}, and J.~{Lal}: Physica
  B Condensed Matter {\bfseries 356} (2005) 223.

\bibitem{2005Bleuel2}
M.~{Bleuel}, K.~{Littrell}, R.~{G{\"a}hler}, and J.~{Lal}: Physica B Condensed
  Matter {\bfseries 356} (2005) 213.

\bibitem{2014Holderer}
O.~Holderer, H.~Frielinghaus, S.~Wellert, F.~Lipfert, M.~Monkenbusch, R.~von
  Klitzing, and D.~Richter: Journal of Physics: Conference Series {\bfseries
  528} (2014) 012025.

\bibitem{1985Boucher}
J.~P. Boucher, F.~Mezei, L.~P. Regnault, and J.~P. Renard: Physical Review
  Letters {\bfseries 55} (1985) 1778.

\bibitem{2002Mezei}
Neutron Spin Echo Spectroscopy -- Basics, Trends and Applications, In F.~Mezei,
  C.~Pappas, and T.~Gutberlet (eds), {\em Scattering}. Springer, Berlin,
  Heidelberg, 2002.

\bibitem{2007Hayashida}
H.~Hayashida, M.~Kitaguchi, M.~Hino, Y.~Kawabata, and T.~Ebisawa: Physica B:
  Condensed Matter {\bfseries 397} (2007) 144.

\bibitem{nGem}
Detectors (nGem \& Thin-Gem) are commercially available from BeeBeans
  Technologies, Japan. For details see: https://www.bbtech.co.jp.

\bibitem{2014Kindervater}
J.~Kindervater, W.~H\"au\ss{}ler, M.~Janoschek, C.~Pfleiderer, P.~B\"oni, and
  M.~Garst: Phys. Rev. B {\bfseries 89} (2014) 180408.

\bibitem{2013Weber}
T.~Weber, G.~Brandl, R.~Georgii, W.~H{\"a}u{\ss}ler, S.~Weichselbaumer, and
  P.~B{\"o}ni: Nuclear Instruments and Methods in Physics Research Section A:
  Accelerators, Spectrometers, Detectors and Associated Equipment {\bfseries
  713} (2013) 71.

\bibitem{2011Brandl}
G.~Brandl, R.~Georgii, W.~H{\"a}ussler, S.~M{\"u}hlbauer, and P.~B{\"o}ni:
  Nuclear Instruments and Methods in Physics Research Section A: Accelerators,
  Spectrometers, Detectors and Associated Equipment {\bfseries 654} (2011) 394.

\bibitem{2018Martin}
N.~Martin: Nuclear Instruments and Methods in Physics Research Section A:
  Accelerators, Spectrometers, Detectors and Associated Equipment {\bfseries
  882} (2018) 11.

\bibitem{2016Oda}
T.~Oda, M.~Hino, M.~Kitaguchi, P.~Geltenbort, and Y.~Kawabata: Review of
  Scientific Instruments {\bfseries 87} (2016) 105124.

\bibitem{2017Hino}
M.~Hino, T.~Oda, N.~L. Yamada, H.~Endo, H.~Seto, M.~Kitaguchi, M.~Harada, and
  Y.~Kawabata: Journal of Nuclear Science and Technology {\bfseries 54} (2017)
  1223.

\bibitem{2017Nakajima}
K.~Nakajima, Y.~Kawakita, S.~Itoh, J.~Abe, K.~Aizawa, H.~Aoki, H.~Endo,
  M.~Fujita, K.~Funakoshi, W.~Gong, M.~Harada, S.~Harjo, T.~Hattori, M.~Hino,
  T.~Honda, A.~Hoshikawa, K.~Ikeda, T.~Ino, T.~Ishigaki, Y.~Ishikawa, H.~Iwase,
  T.~Kai, R.~Kajimoto, T.~Kamiyama, N.~Kaneko, D.~Kawana, S.~Ohira-Kawamura,
  T.~Kawasaki, A.~Kimura, R.~Kiyanagi, K.~Kojima, K.~Kusaka, S.~Lee,
  S.~Machida, T.~Masuda, K.~Mishima, K.~Mitamura, M.~Nakamura, S.~Nakamura,
  A.~Nakao, T.~Oda, T.~Ohhara, K.~Ohishi, H.~Ohshita, K.~Oikawa, T.~Otomo,
  A.~Sano-Furukawa, K.~Shibata, T.~Shinohara, K.~Soyama, J.-i. Suzuki,
  K.~Suzuya, A.~Takahara, S.-i. Takata, M.~Takeda, Y.~Toh, S.~Torii,
  N.~Torikai, N.~L. Yamada, T.~Yamada, D.~Yamazaki, T.~Yokoo, M.~Yonemura, and
  H.~Yoshizawa: Quantum Beam Science {\bfseries 1} (2017) 9.

\bibitem{2011Georgii}
R.~Georgii, G.~Brandl, N.~Arend, W.~Häußler, A.~Tischendorf, C.~Pfleiderer,
  P.~Böni, and J.~Lal: Applied Physics Letters {\bfseries 98} (2011) 073505.

\bibitem{2003Haussler}
W.~H{\"a}u{\ss}ler, U.~Schmidt, G.~Ehlers, and F.~Mezei: Chemical Physics
  {\bfseries 292} (2003) 501.

\bibitem{2018Franz}
C. Franz, O. Soltwedel, S. S\"aubert, A. Wendl, W. Gottwald, F. Haslbeck, L.
  Spitz, C. Pfleiderer, Longitudinal Neutron Resonance Spin Echo Spectroscopy
  under Large Energy Transfers, Proceeding of PNCMI 2018, Journal of Physics:
  Conference Series, accepted for publication (2019).

\bibitem{2018Wendl}
A.~Wendl: Masterthesis, Technische Universiti\"at M\"unichen, Garching (2018).

\bibitem{Wendl-preprint}
A. Wendl, S. S\"aubert, C. Duvinage, J. Kindervater, P. Dharmalingam, A.
  Boothroyd, C. Franz, C. Pfleiderer, MIEZE spectroscopy of spin dynamics and
  crystal field excitations in {Ho$_2$Ti$_2$O$_7$}, unpublished (2019).

\bibitem{2003Bleuel}
M.~Bleuel: {PhD}, Technische Universit\"at M\"unchen, Garching (2003).

\bibitem{2019Jochum}
J. K. Jochum, A. Wendl, T. Keller, C. Franz, Influence of the current of the
  field reduction coil on the MIEZE contrast, unpublished (2019).

\bibitem{2011Haeussler}
W.~H{\"a}ussler, P.~B{\"o}ni, M.~Klein, C.~J. Schmidt, U.~Schmidt, F.~Groitl,
  and J.~Kindervater: Review of Scientific Instruments {\bfseries 82} (2011)
  045101.

\bibitem{2016Koehli}
M.~K{\"o}hli, M.~Klein, F.~Allmendinger, A.-K. Perrevoort, T.~Schr{\"o}der,
  N.~Martin, C.~J. Schmidt, and U.~Schmidt: Journal of Physics: Conference
  Series {\bfseries 746} (2016) 012003.

\bibitem{2001Bramwell}
S.~T. Bramwell: Science {\bfseries 294} (2001) 1495.

\bibitem{2009Fennell}
T.~Fennell, P.~P. Deen, A.~R. Wildes, K.~Schmalzl, D.~Prabhakaran, A.~T.
  Boothroyd, R.~J. Aldus, D.~F. McMorrow, and S.~T. Bramwell: Science
  {\bfseries 326} (2009) 415.

\bibitem{2012Krey}
C.~Krey, S.~Legl, S.~R. Dunsiger, M.~Meven, J.~S. Gardner, J.~M. Roper, and
  C.~Pfleiderer: Phys. Rev. Lett. {\bfseries 108} (2012) 108.257204.

\bibitem{2004Ehlers}
G.~Ehlers, A.~L. Cornelius, T.~Fennell, M.~Koza, S.~T. Bramwell, and J.~S.
  Gardner: Journal of Physics: Condensed Matter {\bfseries 16} (2004) S635.

\bibitem{2009Clancy}
J.~P. Clancy, J.~P.~C. Ruff, S.~R. Dunsiger, Y.~Zhao, H.~A. Dabkowska, J.~S.
  Gardner, Y.~Qiu, J.~R.~D. Copley, T.~Jenkins, and B.~D. Gaulin: Physical
  Review B {\bfseries 79} (2009) 014408.

\bibitem{2017Ruminy}
M.~Ruminy, S.~Chi, S.~Calder, and T.~Fennell: Phys. Rev. B {\bfseries 95}
  (2017) 060414.

\bibitem{muhlbauer}
S. M{\"u}hlbauer, Diploma thesis, Technical University of Munich (2005).

\bibitem{2012Holmes}
A.~T. Holmes, G.~R. Walsh, E.~Blackburn, E.~M. Forgan, and M.~{Savey-Bennett}:
  Review of Scientific Instruments {\bfseries 83} (2012) 023904.

\bibitem{2013Liebi}
M.~Liebi, P.~G. van Rhee, P.~C.~M. Christianen, J.~Kohlbrecher, P.~Fischer,
  P.~Walde, and E.~J. Windhab: Langmuir {\bfseries 29} (2013) 3467.

\bibitem{2019_Saeubert_PRB}
S.~S\"aubert, J.~Kindervater, F.~Haslbeck, C.~Franz, M.~Skoulatos, and
  P.~B\"oni: Phys. Rev. B {\bfseries 99} (2019) 184423.

\bibitem{Kindervater-PhD}
J.~Kindervater: {{PhD Thesis}}, Technische Universit\"at M\"unchen (TUM)
  (2015).

\bibitem{1940Holstein}
T.~Holstein and H.~Primakoff: Physical Review {\bfseries 58} (1940) 1098.

\bibitem{1966Keffer}
F.~Keffer, Spin Waves, In S.~Fl\"ugge and H.~P.~J. Wijn (eds), {\em
  Ferromagnetism / {{Ferromagnetismus}}}, Vol. 4 / 18 / 2, p.~1. {Springer
  Berlin Heidelberg}, 1966.

\bibitem{1969Collins}
M.~F. Collins, V.~J. Minkiewicz, R.~Nathans, L.~Passell, and G.~Shirane:
  Physical Review {\bfseries 179} (1969) 417.

\bibitem{Soltwedel-preprint}
O. Soltwedel, L.Spitz, J. K. Jochum, A. Wendl, C. Pfleiderer, C. Franz, MIEZE
  spectroscopy of sub-picosecond collective bulk dynamics in bulk liquid water,
  unpublished (2019).

\bibitem{2016:Arbe:PRL}
A.~Arbe, P.~Malo~de Molina, F.~Alvarez, B.~Frick, and J.~Colmenero: Phys. Rev.
  Lett. {\bfseries 117} (2016) 185501.

\bibitem{2012:Monkenbusch:IFF}
M.~Monkenbusch, Instruments for neutron scattering, In M.~Angst, T.~Br\"uckel,
  D.~Richter, and R.~Zorn (eds), {\em 43$^\mathrm{rd}$ IFF Spring School:
  Scattering Methods for Condensed Matter Research: towards novel applications
  at future sources}. Forschungszentrum J\"ulich, D\"uren, 2012.

\bibitem{2019_Haslbeck_PRB}
F.~Haslbeck, S.~S\"aubert, M.~Seifert, C.~Franz, M.~Schulz, A.~Heinemann,
  T.~Keller, P.~Das, J.~D. Thompson, E.~D. Bauer, C.~Pfleiderer, and
  M.~Janoschek: Phys. Rev. B {\bfseries 99} (2019) 014429.

\end{thebibliography}

\end{document}